\documentstyle[12pt]{article}
\begin{document}
\vskip 4 cm
\begin{center}
\Large{\bf A Classification of the Fundamental 3+1+1 Forces}
\end{center}
\vskip 3 cm
\begin{center}
{\bf S. Afsar Abbas} \\
Centre for Theoretical Physics\\
JMI, New Delhi - 110025, India\\
(e-mail : afsar.ctp@jmi.ac.in)
\end{center}
\vskip 15 mm  
\begin{centerline}
{\bf Abstract }
\end{centerline}
\vskip 3 mm

The Dark Energy problem is forcing us to re-examine our 
models and our understanding of relativity and space-time. 
The Standard Model of particle physics 
and its extensions are already in
crisis. Having failed so far to include gravity in a 
proper unified framework, these are now faced with an additional 
unwanted fifth force of repulsion. How does one understand
this 3+1+1 fundamental force dilemma? Quite clearly this 
points to a limitation of our present understanding and demands 
extension of our theoretical framework. 
To be able to go beyond these limitations, here we
introduce a novel idea of the Fundamental Forces. 
This allows us to perceive the General Theory of Relativity 
and Einstein's Equation from a different perspective. 
This will give us an additional and an all-encompassing way of 
classifying these five fundamental forces in a consistent manner.
In addition to providing us with an improved understanding of 
space and time, it will be shown how it leads to a 
resolution of the Dark Energy problem.

\newpage

Dark Energy is certainly the most puzzling problem in physics and 
astronomy today [1]. All kind of proposals,
mostly ad-hoc in nature, to solve the problem, 
are being put forward. But we are nowhere near a resolution of 
the issues involved. 

So far all our understanding of nature has been successfully
described within the Standard Model (SM) of particle physics.
Whatever was not accessible to it, has been explained in terms of various
theoretical extensions of the SM.
All this was done in terms of an understanding that there are four
fundamental forces. Three of these are gauge forces and the fourth one, 
that of gravity, it is believed, shall "soon" be incorporated in a
unified whole as some kind of quantized gauge theory. 
This "soon" has been dogging us for several decades.
The problem becomes more confusing in that there always remains a 
clear possibility that gravity, at a fundamental level, may 
be a different kind of force altogether and may not be quantized
at all, and in which case its unification with the other three forces 
will have to be seen differently.
The fact that one has not been able to achieve this so called unification
of the four forces so far, we
are thus justified in breaking this so called four force problem as 
actually being of the nature of a 3+1 force problem. 

Given the above situation,
no one expected and no one wanted, yet another new 
fundamental "force" to spring up.
But there it is - the new force of repulsion of galaxies [1],
call it RF (Repulsive Force)! 

One question that arises immediately is, as 
to the nature of this RF. Is it a simply a gauge force like the other 
three and then the force problem is of the 4+1 kind; or 
is it fundamentally of the gravity kind and in which case the force 
problem is that of 3+2 kind; or is it different from all these 
and in which case it is 3+1+1 kind? 

To understand this, let us look at the Einstein's Equation.
Harvey and Schucking [2] correcting for Einstein's error 
in understanding the role of the cosmological term $\lambda$,
have derived the most general equation of motion to be

\begin{equation}
G_{\mu \nu} + \lambda {g_{\mu \nu}} 
= 8 \pi G \langle \phi | T _{\mu \nu} | \phi \rangle
\end{equation}
 
They showed that the Cosmological Constant $\lambda$ 
above provides 
a new repulsive force proportional to mass m, repelling every 
particle of mass m with a force

\begin{equation}
F = m {c^2} {\lambda \over 3} x
\end{equation}

Recent data [1] on $\lambda$ is what leads to the crisis of 
Dark Energy.

The situation is akin to the discovery of the muon, when people were 
quite happy and contended with only the electron and when I. I. Rabi 
in puzzlement asked, "who ordered it?" 
We too can paraphrase Rabi by asking, "Who ordered this fifth force?" 
The discovery of muon forced scientists to extend their theoretical 
framework significantly.
No patch-up work, but a genuine attempt to include this new force
in a fundamental and consistent framework of our understanding of 
nature.  

It may be remarked that the concept of a so called  
fifth force has been there for quite 
sometime. Extensions of Einstein's GTR, like for example Brans-Dicke 
theory, necessarily have an extra fifth force, in which case the RF may 
belong to the 3+2 or 3+1+1 classification. 
Higher dimentional
Kaluza-Klein kind of theories, supersymmetric theories, superstring 
theories etc also predict the fifth fundamental force of the Yukawa kind 
and in which case it will very likely belong to the 4+1 kind. It is not 
clear that the new RF is this putative fifth force [1,3].
In fact this theoretical fifth force is incompatible with overall
cosmological framework [1,3].
Just because the word "fifth" force has been usurped by the other
models, does not mean that the actual empirical fifth RF is of their kind.
So minimal conclusion would be that with the new RF, the force problem is 
per se of the 3+1+1 kind. 

Here we wish to understand the "force" nature of the new problem.
To do so we introduce a 
new concept of the "Universal Force". It was first proposed 
by Hans Reichenbach [4]. It is a genuine scientific concept, which having
been proposed in a book called "philosophical", has unfortunately
not been accessible to physicists by and large.
Reichenbach's "lost" work on the three-valued logic for quantum 
mechanics, in recent years, has found its way in physics literature.
The concept of the "universal Force" deserves it, actually more so!
Rudolf Carnap in the Introduction of 
Reichenbach's book [4] called the concept of the 
Universal Forces, " ... of great 
interest for the methodology of physics but what has so far 
not received the attention it deserves".
In this paper we shall try to rectify for this failure of 
appreciating the concept of the Universal Forces - albeit
in a somewhat altered and improved manner.

Reichenbach defines two kind of forces - Differential Forces and 
Universal Forces. It may be pointed out that 
the term "force" here should not be taken strictly as
defined in physics but in a broad and general framework. 
In fact Carnap has suggested that the term "effect" 
instead of "force' would better serve the purpose [4]
and which allows it be used in different frameworks. Hence to 
conform with the accepted practice, though in this paper we 
shall continue to use the term "Universal Force" the reader may do 
well to remember that what we really mean is "Universal Effect".

One calls a force Differential if it acts differently on different 
substances. It is called Universal if it is quantitatively the 
same for all the substances [4,5]. If we heat a rod of initial 
length $l_0$ from initial temperature $T_0$ to temperature T then 
its length is given as

\begin{equation}
l = l_0 [ 1 + \beta ( T - T_0 ) ]
\end{equation}

where $\beta$ the coefficient for thermal expansion is different 
for different materials. Hence this is a Differential Force.
Now the correction factor due to the influence of gravitation on 
the length of the rod is

\begin{equation}
l = l_0 [ 1 - C { m \over r } {cos{^2} \phi}]
\end{equation}

Here the rod is placed at a distance r from sun whose mass is m 
and $\phi$ is the angle of the rod with respect to the the line 
sun to rod. C is a universal constant ( in CGS unit 
C= 3.7 x ${10}^{-29}$ ). As this acts in the same manner for any 
material of mass m, gravity is a Universal Force
as per the above definition.

Reichenbach also gives a general definition of the Universal 
Forces [4,p 12] as: (1) affecting all the materials in the same 
manner and (2) there are no insulating walls against it. We saw 
above that gravity is such a force,

Indeed gravity is a Universal Force par excellence. It affects all 
matter in the same manner. The equality of the gravitational and 
inertial masses is what ensures this physically. If the 
gravitational and inertial masses were not found to be equal, then 
one would not have been able to visualize of the paths of freely 
falling mass points as geodesics in the four dimentional space-time.
In that case different geodesics would have resulted from 
different materials of mass points [4]. 

Therefore the universal effect of gravitation on different kinds 
of measuring instruments is to define a single geometry for all of 
them. Viewed this way, one may say that gravity is 
geometerized. "It is not theory of gravitation that becomes 
geometry, but it is geometry that becomes the experience of the 
gravitational field" [4, p 256]. Why does the planet follow the 
curved path? Not because it is acted upon by a force but 
because the curved space-time manifold leaves it with no other 
choice!

So as per Einstein's theory of relativity, one does not speak of 
a change produced by the gravitational field in the measuring 
instruments, but regard the measuring instruments as free from any 
deforming forces. Gravity being a Universal Force, in the 
Einstein's Theory of Relativity, it basically disappears and is 
replaced by geometry.

In fact Reichenbach [4, p 22] shows how one can give a consistent 
definition of a rigid rod - the same rigid rods which are needed 
in relativity to measure all lengths. "Rigid rods are solid bodies 
which are not affected by Differential Forces, or concerning which 
the influence of Differential Forces has been eliminated by 
corrections; Universal  Forces are disregarded. We do not neglect 
Universal Forces. We set them to zero by definition. Without such 
a rule a rigid body cannot be defined." In fact this rule also 
helps in defining a closed system as well. 

All this was formalized in terms of a theorem by 
Reichenbach [4, p 33]

\vskip 1 cm

{\bf THEOREM $\theta$} :

Given the geometry $G^0$ 
to which the measuring instruments 
conform, we can imagine a Universal Force F which affects 
the instruments in such a way that the actual geometry is an 
arbitrary geometry $G$, while the observed deviation from $G$
is due to universal deformation of the measuring instruments."

\begin{equation}
{G^0} + F = G 
\end{equation}

Hence only the combination ${G^0} + F$ is testable. 
As per Reichenbach's 
principle one prefers the theory wherein we put F=0.
If we accept Reichenbach principle of putting the 
Universal Force of gravity to zero, then the arbitrariness in the 
choice of the
measuring procedure is avoided and the question of the geometrical 
structure of the physical space has a unique answer determined by 
physical measurement. It is this principle which Carnap praises 
highly [5, p 171], " Whenever there is a system of physics in 
which a certain universal effect is asserted by a law that 
specifies 
under what conditions in what amount the effect occurs, then the 
theory should be transformed so that the amount of effect would be 
reduced to zero. This is what Einstein did in regard to 
contraction and expansion of bodies in gravitational field."
The left hand side of Einstein's equation (below)
gives the relevant non-Euclidean geometry

\begin{equation}
G_{\mu \nu} = 8 \pi G \langle \phi | T _{\mu \nu} | \phi \rangle
\end{equation}

(Note that we suppress $\lambda$ here).
In the case of gravity, and in as much as Einstein's Theory of 
Relativity has been well tested experimentally, we treat the 
above concept as well placed empirically. But from this single 
success Reichenbach generalizes this as a fundamental principle 
for all cases where Universal forces may arise. As Carnap states 
[5, p 171], " Whenever universal effects are found in physics, 
Reichenbach maintained that it is always possible to eliminate 
them by suitable transformation of theory; such  
a transformation should be made because of the overall simplicity 
that would result. This is a useful general principle, 
deserving more attention than it has received.
It applies not only to relativity theory, but also to situations 
that may arise in the future in which other universal effects may 
be observed. Without the adoption of this rule there is no way to 
give unique answer to the question - 
what is the structure of space?". 

As such Reichenbach goes ahead and tries to apply this principle 
of elimination of Universal Forces to another universal effect 
that he finds and which arises from considerations of 
topology ( as an additional consideration over and above that of 
geometry ) of space-time of the universe.

The Theorem $\theta$ is limited to talking about the geometry of
space-time only. It does not take account of specific 
topological issues 
that may arise. To take account of topology of the space-time 
we shall have to extend the said theorem appropriately.

What would one experience if space had different topological 
properties. To make the point home Reichenbach considers a 
torus-space [4, p 63]. This is quite detailed and extensive.
However for the purpose of simplifying the
and shortening the discussion here we shall
talk of a two dimensional being who lives on the 
surface of a sphere. His measurements tell him so. But in
spite of this he insists that he lives on a plane. 
He may actually do so as per our discussion above if he confines 
himself to metrical relations only.
With an appropriate Universal Force he can he can justify living
on a plane. But the surface of a sphere is topologically 
different 
from that of a plane. On a sphere if he starts at a point X and 
goes on a world tour he may come back to the same point X. But 
this is impossible on a plane. And hence 
to account for coming back to the "same point" 
he has to maintain that on the plane he 
actually has come back to a different point Y - which though is 
identical to X in all other respects. 
One option for him is to 
accept that he is actually living on a sphere. 
However if he still wants 
to maintain his position that he is living on a plane then he has 
to explain as to how point Y is 
physically identical to point X in spite of 
the fact that X and Y are different and distinct points of space. 
Indeed he can do so by visualizing a fictitious force as an 
effect of some kind of "pre-established harmony" [4, p 65] by 
proposing that everything that occurs at X also occurs at the 
point Y. As it would affect all matter in the same manner this 
corresponds to a Universal Force/Effect as per Reichenbach's 
definition.

This interdependence of corresponding points which is essential 
in this "pre-established" harmony cannot be interpreted as 
ordinary causality, as it does not require ordinary time to 
transmit it
and also does not spread continuously through intervening space. 
Hence there is no mysterious causal connection between the points 
X and point Y. Thus this necessarily entails 
proposing a "causal anomaly" [4, p 65].
In short connecting different topologies through a fictitious
Universal Effect of "pre-established harmony" necessarily calls 
for introduction of "causal anomalies". 
Call this new hypothesize 
Universal Force as A and the Theorem $\theta$ be extended to 
read

\begin{equation}
{G^0} + F  + A = {\bf G} 
\end{equation}

where on the right had side we have given a different 
capital ${\bf G}$ 
which reduces to $G$ of the original Theorem $\theta$
when A is set equal to zero.

Now as per Reichenbach's law of preferring that physical reality 
wherein all Universal Forces are put to zero, he 
advocates of putting A to zero. He pointed out that this has the 
advantage of retaining physical "causality " in our science. 

However, as the said 'causal anomaly" is of topological origin we 
cannot be sure in what manner it will manifest itself physically. 
In addition
will not the Universal Force/Effect of "pre-established harmony"
compensate for it in some manner?
So what one is saying is that it is possible that Reichenbach was 
wrong in putting all Universal Forces to zero. 
It was fine to put F to zero, which allowed us to define a truly
"rigid" rod and which let to a geometrical interpretation of 
gravity in a unique manner.
But in the case of this new topological Universal Force we 
really do not know enough 
and let us not be governed by any theoretical prejudice and 
let the Nature decide as to what is happening. So to say, let us 
look at modern cosmology to see if it is throwing up any new 
Universal Forces which may be identified with our 
"pre-established harmony" here.

Not known to Reichenbach at his time, but now known to us, and as 
discussed above, there is indeed another "Universal Force"
of repulsion RF. It is universal as it acts in the same manner on all
bodies of mass m. This new fifth fundamental force, which is a
puzzle for the SM and its putative extensions, is but a natural ally of 
gravity in being of universal character.

Quite clearly this repulsive force
is a new Universal Force as per our definition
and hence conforms to the "pre-established harmony" aspect of the 
"causal anomaly".
Thus we see that indeed as per the recent data on accelerating 
universe we have stumbled upon this new Universal Force which is 
of topological origin. Hence the source of dark energy is 
due to "causal anomaly" arising from the unique topological 
structure of our universe.
This solves the mystery of the origin of Dark Energy.

So we would like to emphasize that it is the accelerating 
universe ( and hence the Dark Energy ) 
which is forcing us to accept the incorporation of this
"causal anomaly" of topological origin.  
Implications of this new concept in physics have now to be 
explored.

We know that the surface's values which do not change with deformation are 
called "topology" of the surface. A surface's "geometry" consists of those 
properties which do change with deformation of the surface.
Viewed this way, topology and geometry are complementary properties of 
space-time. Indeed this is how gravity ( related to geometry ) and the new 
repulsive-force ( related to topology) arise as two fundamental
and complementary Universal Forces.

Why one is attractive and the other repulsive, is something that may be 
part of this complementarity. We do not understand that yet, but
the very fact that they are both of universal character should
allow us to understand them better in future.  

So as per this new classification, there are three well known gauge 
forces and two universal forces - that of gravity and the new one
of repulsion. 
However, this has an advantage that it points to a basic
similarity between the two - gravity and repulsive-force, 
which is not apparent in the canonical way of 
adding up the fifth force in an ad-hoc manner.
Hence as per the definition above, the forces should be 
classified as 3+2 kind. Clearly this is providing us with an understanding 
which may help us in the present puzzling scenario.
It is allowing us to understand the nature of this new RF without
contradicting anything known today.

One would like to ask as to in what other manner, incorporation of 
this new "causal anomaly", may help us in understanding Nature 
better? Will it provide new perspectives as 
answers to quantum mechanical puzzles of 
quantum jumps, non-locality etc. These are open questions  
to be tackled in future. 

In the early twentieth century, one knew of only two fundamental forces
- the gravitational and the electromagnetic. Having obtained
his equation of motion in GTR, Einstein tried to unify electromagnetism
within the same geometrical framework. Others like Weyl also tried to do 
so. Kaluza's higher dimentional idea was aimed at the same target. With 
hindsight we know now that they failed because they did not take account 
of the strong and the weak forces. We have also learned that these two 
along with electromagnetic are gauge forces. And also as discussed above,
gravity has not fallen in line as of now. Could it be that we have failed 
to get a Theory of Everything so far, as like earlier, we were not 
aware of all the forces. 
We should re-examine our understanding of what we mean by 
force, expand out in proper direction an see if there are other ways of 
looking at a "force" different from the canonical manner of looking at it
and which we teach to our undergraduates. 
Quite clearly we have no clue as to why there are five (3+1+1) forces?
Can we be sure that this is it and no more fundamental forces will make 
their presence felt in the future. If they do, then present accepted point 
of view will be at a loss to account for it. But the UF idea presented 
above will naturally incorporate it. This is the power of the UF idea.
In this paper we have pointed
out a very fruitful approach which not only allows us to look at the new
RF in a new manner, but also points to a new a direction on Dark Energy
as well.   

Niels Bohr always rued that a proper understanding of quantum mechanics
was too much shackled by language.
Perhaps he was right. The shackles put by the canonical way of looking
at the new unwanted fifth force may be the cause of present confusion.
The new framework of Universal Forces is allowing us to expand our 
vocabulary and understanding and is hopefully making us more competent 
in tackling the new puzzles of the fifth force.

\newpage

\vskip 1 cm
\begin{center}
{\bf REFERENCES }
\end{center}

\vskip 1 cm

1. E J Copeland, M Sami and S Tsujikawa, "Dynamics of Dark Energy"
{\it Int J Mod Phys}, {\bf D 15} (2006) 1753-1936

\vskip 1 cm

2. A Harvey and E Schucking, "Einstein's mistake and the 
cosmological constant", 
{\it Am J Phys}, {\bf 68} (2000) 723-727

\vskip 1 cm

3. E I Guendelmann and A B Kaganowich, "Dark Energy and the Fifth Force 
Problem", {\it J Phys}, {\bf A 41} (2008) 164053

\vskip 1 cm

4. H Reichenbach, "The philosophy of space and time", Dover, 
New York (1957) (Original German edition in 1928)

\vskip 1 cm

5. R Carnap, "An introduction to the philosophy of science",
Basic Books, New York (1966)

\end{document}